\documentclass{elsart}
\usepackage{adnd}
\usepackage{longtable}
\usepackage{mathptmx}
\usepackage[square,sort&compress]{natbib}
\begin{document}

\begin{frontmatter}

\journal{Atomic Data and Nuclear Data Tables}

\title{Compilation of Giant Electric Dipole Resonances Built on Excited States}

\author{A. Schiller\corauthref{AS}}
\corauth[AS]{Corresponding author: NSCL/MSU, 1 Cyclotron, East Lansing, MI 
48824\@. Phone: (517) 324-8142, Fax: (517) 353-5967}
\address{National Superconducting Cyclotron Laboratory, Michigan State 
University, East Lansing, MI 48824}
\ead{schiller@nscl.msu.edu}

\and

\author{M. Thoennessen}
\address{National Superconducting Cyclotron Laboratory, and \\
Department of Physics \& Astronomy, Michigan State University, East Lansing, MI
48824}

\date{5.5.2006}

\begin{abstract}
Giant Electric Dipole Resonance (GDR) parameters for $\gamma$ decay to excited
states with finite spin and temperature are compiled. Over 100 original works 
have been reviewed and from some 70 of which more than 300 parameter sets of 
hot GDR parameters for different isotopes, excitation energies, and spin 
regions have been extracted. All parameter sets have been brought onto a common
footing by calculating the equivalent Lorentzian parameters. The current 
compilation is complementary to an earlier compilation by Samuel S. Dietrich 
and Barry L. Berman (At.\ Data Nucl.\ Data Tables \bf 38\rm, 199-338, 1988) on 
ground-state photo-neutron and photo-absorption cross sections and their 
Lorentzian parameters. A comparison of the two may help shed light on the 
evolution of GDR parameters with temperature and spin. The present compilation 
is current as of January 2006.
\end{abstract}

\end{frontmatter}

\newpage

\tableofcontents

\listofDtables

\section{INTRODUCTION}

Over the last decades, numerous data on hot Giant Electric Dipole Resonance 
(GDR) parameters have been accumulated. Since the field seems to have matured, 
it might be useful at this point to gather all present data in one 
comprehensive compilation in a uniform format. The Introduction is organized in
the following way: first, we give some theoretical motivation why GDR 
parameters might be dependent on temperature, hence what question the present 
compilation tries to address. Then, in Subsection \ref{sect:exp}, we give an 
overview over typical experimental techniques used for measuring hot GDR 
parameters; in Subsection \ref{sect:smc} we describe different ways of how hot 
GDR parameters are extracted from such experiments by means of 
statistical-model calculations. Although we have not attempted to extract any 
GDR parameters from such calculations ourselves, we believe it is helpful for 
the reader to get a general impression of the experimental and data-analysis 
part of the compiled works. In Subsection \ref{sect:data} we explain how 
different sets of GDR parameterizations from the original articles were brought
onto a common footing. Finally, we end the Introduction with a statement 
regarding our policy and a note on references.

\subsection{Theoretical considerations}

A first good understanding of statistical $\gamma$ emission was gained from the
works of Brink \cite{Br55} and Axel \cite{Ax62} who realized that average 
electric-dipole ($E1$) transition strengths in different energy regimes can be 
described in a unified fashion by assuming that the GDR can be built on any 
excited state, and that the GDR properties do not depend on the temperature of 
the excited state in question. 

This so-called Brink-Axel hypothesis has been refined in the past to allow for 
temperature- and spin-dependent widths. A model to motivate such a modification
takes into account shape fluctuations of the nucleus. Since the ground-state 
GDR splits into two components for a nucleus with static deformation, and the 
splitting depends on the degree of deformation \cite{AF98}, it is reasonable to
assume that at finite temperatures, when the nucleus can explore a large volume
in deformation space, the GDR response will be an average over different 
deformations and hence, different splittings. The result within this adiabatic
damping model will be a more diffuse and certainly wider GDR than the 
ground-state GDR\@.

Quantitatively, assuming a Fermi-gas level density, one can write the nuclear 
entropy in the microcanonical ensemble as $S=2\sqrt{a(E-V)}$, where $V$ is some
potential energy proportional to the square of the deformation $V=k\beta^2$\@. 
The entropy is trivially maximized for $\beta=0$; expanding $S$ for small 
$V\ll E$ yields $S\approx 2\sqrt{aE}\left(1-\frac{V}{2E}\right)$\@. With 
$T=\sqrt{E/a}$, the probability distribution to find a nucleus with energy $E$ 
and deformation $\beta$ becomes 
$P\approx\exp\left(2\sqrt{aE}\right)\,\exp\left(-k\beta^2/T\right)$ where the 
second factor represents a Gaussian distribution of deformations around 
$\beta=0$ and a width of $\sqrt{T/2k}$\@. Assuming that the splitting of the 
GDR into two components is roughly proportional to the nuclear deformation, the
shape-fluctuation model predicts an increase in width of the GDR roughly 
proportional to $\sqrt{T}$\@. Moreover, there is also a potential spin 
dependence of the GDR width which stems from the possibility of spin-induced 
deformation. Finally, orientation fluctuations of the nucleus and 
nuclear-structure effects such as pairing can influence the temperature 
dependence of the GDR in different energy regimes. Several groups have 
calculated temperature-dependent GDR widths along these lines 
\cite{KA98,OB96,AD01} and a simple scaling law has emerged \cite{KA98}\@.

Investigations of the low-energy tail of the GDR have also yielded indications 
for a temperature-dependent GDR width. It was, e.g., noted by Popov \cite{Po82}
in $(n,\gamma\alpha)$ experiments on Sm nuclei that the $\gamma$ strength 
function tends to approach a finite value for $E_\gamma\rightarrow 0$ for 
$\gamma$ transitions in the quasicontinuum (below the neutron separation 
energy)\@. This experimental observation led Kadmensk\u{\i}i, Markushev, and 
Furman (KMF) to propose a $\gamma$ strength-function model for spherical nuclei
with a temperature-dependent width \cite{KM83} based on the effect of in-medium
nucleon-nucleon collisions. The proposed temperature dependence was derived 
within Migdal's theory of Fermi liquids and has the form 
$\Gamma(E_\gamma,T)\propto\left(E_\gamma^2+4\pi^2T^2\right)$\@. The model was 
later improved by Sirotkin \cite{Si86} who included the Pauli exclusion 
principle, and it was extended to deformed nuclei within the framework of the 
generalized Lorentzian model by Kopecky and Uhl \cite{KU90} and by inclusion of
a coupling term between the $E1$ operator and the quadrupole deformation 
according to Mughabghab and Dunford \cite{MD00}\@. The KMF model (taken at 
constant temperatures) and its extensions have been successfully applied to 
improve $\gamma$ and isomeric production cross sections \cite{Gr99} and they 
have been used for direct fits of measured low-energy $\gamma$ strength 
functions \cite{VG01}\@. The connection of collisional-damping models with hot 
GDR parameters has been made in \cite{BC96}\@.

\subsection{Experimental Techniques}
\label{sect:exp}

Unlike the measurement of the GDR by ground-state photo-absorption 
cross-section measurements \cite{DB88}, measurements of hot GDR parameters can 
be performed in many different ways. One of the simplest ways is by 
fusion-evaporation reactions where only $\gamma$ rays are detected 
\cite{FS93}\@. Such measurements are the most inclusive reactions, since the 
high-energy $\gamma$ yield which competes with particle and especially neutron 
evaporation is representative for a range of different product nuclei, 
excitation energies, and spins. Moreover, it is not necessarily guaranteed that
all detected $\gamma$ rays stem from fusion-evaporation reactions. Other 
reactions such as inelastic or deep inelastic scattering can compete and yield 
$\gamma$ rays from target or projectile-like fragments. 

To improve the sensitivity of such experiments to the fusion-evaporation 
reaction channel, typical gates such as, e.g., $\gamma$-multiplicity filters 
\cite{RC03}, detection of heavy evaporation residues \cite{BG89}, and detection
of evaporated, light charged particles such as protons or $\alpha$ particles 
\cite{LG96} can be performed. The resulting $\gamma$-ray spectra are more 
exclusive, not only in terms of the product nuclei from which high-energy 
$\gamma$ rays are emitted, but also in terms of the spin and the 
excitation-energy range investigated. For example, a gate on different $\gamma$
folds translates rather directly into certain spin regions of the investigated 
product nucleus \cite{RC03}\@. A gate on evaporated light charged particles 
will not only reduce the average charge and mass of the product nucleus, but it
will also reduce its average excitation energy, since the evaporated particles 
will carry away some part of the initial excitation energy of the compound 
nucleus \cite{KL97}; hence applying such a gate will test the GDR at somewhat 
lower temperatures than the fully inclusive experiment. In the same way, gating
on the $\gamma$ sum energy \cite{GB88} or on specific product nuclei by means 
of detecting in coincidence discrete, known low-energy $\gamma$ transitions 
\cite{KM01} will also influence the average spin and excitation-energy region 
from which the high-energy $\gamma$ rays are emitted, since one effectively 
biases the competition between high-energy $\gamma$ decay and neutron 
evaporation in one or the other direction. Other, more rarely used gating 
conditions are, e.g., the isomeric $\gamma$ decay by discrete transitions 
\cite{SB89} or the $\alpha$ decay of a product nucleus \cite{CB03}, both of 
which have similar implications for the average spin and excitation-energy 
range from which prompt high-energy $\gamma$ rays are observed.

For heavy nuclei, an added difficulty is the possibility of fission of the 
compound nucleus. Typically, for low spins, the production of an evaporation 
residue dominates while for high spins fission will become the dominant exit 
channel \cite{HB98}\@. Hence, by gating on evaporation residues or fission 
fragments, one effectively selects a spin region from which high-energy 
$\gamma$ emission is observed \cite{BT90}\@. In the case of the fission exit 
channel, one also observes high-energy $\gamma$ emission from the fission 
fragments themselves \cite{DB74}, though typically at significantly higher 
$\gamma$ energies owing to the much lower mass of the fission fragments. 
Another complication is the fact that high-energy $\gamma$ emission can occur 
from the compound nucleus (which is desired), or during the saddle-to-scission 
motion after the nucleus has passed the fission barrier \cite{SD00}\@. Also, 
for the heaviest nuclei investigated, it is not clear whether a true compound 
nucleus forms which is confined by some fission barrier or whether one observes
direct fission or just the formation of a mononucleus. Where we suspect the 
latter as in, e.g., \cite{TG96}, the extracted data do not enter the present 
compilation. 

In some cases, one employs inelastic (see, e.g., \cite{BR98}) or deep inelastic
scattering \cite{EB92,DN82,HB88,HK87,TB91} to excite the target or projectile 
nucleus. By measuring the kinetic energy of at least one of the products, one 
can reconstruct the reaction kinematics event by event and it is possible to 
obtain initial excitation-energy indexed coincident high-energy $\gamma$ 
spectra with only one beam energy. Otherwise, the excited nucleus can be 
treated in the same way as a compound nucleus which is formed in a 
fusion-evaporation reaction.

\subsection{Statistical-model calculations and comparison to experimental data}
\label{sect:smc}

High-energy $\gamma$ spectra are typically analyzed using a statistical-model 
calculation. In the first step, total fusion cross sections and maximum 
$\left(l_0\right)$ or average $\left(\langle I_i\rangle\right)$ angular momenta
are determined. Typically, total fusion cross sections can be verified by 
experiment; maximum angular momenta are calculated by the theory of either 
Winther \cite{Wi95} or Swiatecki \cite{Sw82}\@. Average angular momenta can 
then be determined by $\langle I_i\rangle=\frac{2}{3}l_0$\@. In the next step, 
the decay of the highly excited compound nucleus is modeled. In many cases, 
this simply involves a Hauser-Feshbach-type theory \cite{HF52} into which 
particle and $\gamma$ transmission coefficients (sometimes including 
higher-than-$E1$ multipolarities) as well as nuclear level densities enter. 
Typically, particle transmission coefficients are not discussed in great detail
in the compiled works. The level-density models are either the P\"{u}hlhofer 
model \cite{Pu77} (the default in the statistical-model code CASCADE 
\cite{Pu77}) or the Reisdorf model \cite{Re81}\@. The P\"{u}hlhofer model 
relies on the local Dilg \sl et al.\ \rm parameterization \cite{DS73} for 
excitation energies up to and slightly above the nucleon separation energy, 
while it interpolates then to a regime where the level-density parameter $a$ 
becomes proportional to the nuclear mass number $A$\@. The Reisdorf approach 
builds on the generalized superfluid model by Ignatyuk \sl et al.\ \rm 
\cite{II79}, but it uses a global parameterization for the asymptotic level 
density parameter $a$\@. In one case \cite{BR98}, the level-density model by 
Fineman \sl et al.\ \rm \cite{FB94} is used in the data analysis. 

Statistical-model calculations are often adapted to different experimental 
situations. For light compound nuclei near the $N=Z$ line, an isospin-dependent
formalism is often used \cite{FS93}\@. Also, the Wigner energy \cite{MS66} is 
sometimes included in the level-density parameterization \cite{FS93}\@. For 
large excitation energies, pre-equilibrium emission due to direct and 
semi-direct reaction mechanisms are often taken into account (see, e.g., 
\cite{BB94})\@. Especially the PEQAG2 code \cite{Be89} has been developed for 
this purpose. In the case of fissile compound nuclei, the fission channel and 
the decay of excited fission fragments are modeled as well (see, e.g., 
\cite{HB98})\@. When gating conditions were applied in the experiment, they are
usually reflected in the statistical-model calculation as well, which often 
implies the need to use a Monte-Carlo simulation tool such as in \cite{LG96}\@.
In some cases, also asymmetries $a_2$ of $\gamma$ emission are calculated (see,
e.g., \cite{GB88}), however, experimental asymmetries do not enter into the 
present work (with the exception of their influence on the sign of the quoted 
deformation as in, e.g., \cite{GB88})\@.

Typically, calculated high-energy $\gamma$ spectra are compared to their 
experimental counterparts, and the $\gamma$ transmission coefficients 
(parameterized by one- or two-component Lorentzians multiplied by a factor 
$2\pi\,E_\gamma^3$) are varied until the best fit is obtained. If absolute 
values in the high-energy region are compared, the fit is often normalized to 
the data in an energy region of 3--7~MeV (see, e.g., \cite{KS92}), far below 
the peak of the GDR\@. Statistical uncertainties are usually determined in the 
normal fashion by varying GDR parameters until the quality of the fit 
deteriorates. Systematic uncertainties are more difficult to estimate. A good 
way is, e.g., to perform several fits to the experimental data with different 
level-density parameterizations as in \cite{KS90} or differing sets of other 
input parameters into the statistical-model calculation (say, e.g., those which
describe the dynamic of the fission process such as the nuclear viscosity which
governs the timescale of the saddle-to-scission motion as in \cite{HB98})\@. 
The range of resulting GDR parameters might give a good indication of the size 
of the systematic error. 

Other sources of systematic errors concern the experimental conditions. Some of
the most important problems there involve inefficient neutron-$\gamma$ 
discrimination (often done by time-of-flight techniques as in \cite{BR98}, or 
by simply considering $\gamma$ rays only at backward angles as in \cite{FH91}),
contamination of high-energy $\gamma$ spectra by cosmic rays (which can be 
greatly reduced by coincidence measurements as in \cite{GE86}), target 
impurities, pile-up (see, e.g., \cite{KS87} for a thorough investigation of 
these two effects), and add-back issues. Add back is a technique often used for
an array of small detectors where for high-energy $\gamma$ rays one observes a 
significant amount of (i) Compton scattering from one detector into a 
neighboring one, and (ii) pair production with subsequent annihilation $\gamma$
rays being detected in neighboring detectors. The add-back technique remedies 
this situation by adding the deposited energies in neighboring detectors and 
(rightly) consider such events as stemming from one single $\gamma$ ray (see, 
e.g., \cite{GB88})\@. Pile up is a problem especially for large detectors, 
where two or more coincident $\gamma$ rays hit the same detector and their 
energies add up and are falsely registered as one high-energy $\gamma$ ray 
(see, e.g., the discussion in \cite{KS87})\@. If one applies the add-back 
technique, however, pile up can also occur when two coincident $\gamma$ rays 
hit two neighboring detectors. Obviously, for any given detector array, there 
is an optimal balance between the benefit of applying the add-back technique 
and the possible distortions of the high-energy $\gamma$ spectrum due to pile 
up. Where we found that this balance was not met \cite{MS90}, we have rejected 
the data for the present compilation.

Some other physical background involves nuclear bremsstrahlung which is emitted
in the first moments of the fusion process where individual nucleons of the 
projectile are greatly de-accelerated in the proximity of target nucleons and 
emit high-energy $\gamma$ rays (see, e.g., the discussion in \cite{KS87})\@. 
These $\gamma$ rays can either be modeled (and hence subtracted from the 
high-energy $\gamma$ spectrum which is to be fitted by the statistical-model 
calculation as in \cite{BR98}), or they are simply considered as a source of 
systematic error as in \cite{KS87}\@. Another possible source of high-energy 
$\gamma$ rays is the pre-equilibrium $\gamma$ emission during the formation of 
the compound nucleus which has been investigated by measuring high-energy 
$\gamma$ spectra for (isospin) symmetric and asymmetric reactions as in, e.g., 
\cite{PB03,TR95,FC96}\@. Since in the extreme, such reactions essentially probe
the di-nuclear system and not a compound nucleus (see, e.g., \cite{CR95}), the 
resulting data are not entered into the present compilation. In general, in our
compilations we focus more on low-energy data to avoid complications due to 
non-compound sources of high-energy $\gamma$ rays, hence, data concerning the 
saturation or increase of the GDR width at very high excitation energies such 
as in \cite{YK90,FS94,GB87} are typically omitted.

\subsection{Data treatment}
\label{sect:data}

In our data treatment, the first step was to determine the target isotope from 
the context (for those few articles where it was not stated explicitly)\@. 
Ranges of laboratory energy $E_{\mathrm{lab}}$, initial excitation energy 
$E_{\mathrm{ex}}$, and initial spin $I_i$ were replaced by their central 
values. Average initial spins $\langle I_i\rangle$ were determined from maximum
spins $l_0$ by means of $\langle I_i\rangle=\frac{2}{3}l_0$ for the case of 
fusion-evaporation reactions but irrespective of gating conditions. Ranges in 
final spin $I_f$ after GDR $\gamma$ emission were replaced by central values as
well; the widths of the $I_f$ ranges were converted into $\mathrm{FWHM}$ values
which were preferred in this case over regular uncertainties due to the often 
non-Gaussian distribution of final spins. 

Hot GDR parameterizations can take many different forms. The most common is 
probably the parameterization in terms of a centroid $E$, width $\Gamma$, and 
maximum $\sigma$ of an equivalent Lorentzian photon-absorption cross section. 
The maximum $\sigma$ is often formulated as a fraction $S$ of the 
Thomas-Reiche-Kuhn (TRK) sum rule which describes the integral 
$\frac{\pi}{2}\sigma\Gamma=60\frac{NZ}{A}$~MeV~mb of the Lorentzian in terms of
neutron $N$, proton $Z$, and mass $A$ numbers \cite{DB88}\@. For two-component 
Lorentzian parameterizations, often the total $S_1+S_2$, and either the ration 
$S_2/S_1$ or the relative fraction $F_2=S_2/(S_1+S_2)$ are given. In either 
case, the given parameters were converted into $S_1$ and $S_2$ values. 
Sometimes, the total is assumed to fulfill the TRK while only the ratio or the 
relative fraction $F_2$ is determined by the fit. In these cases, the errors 
are marked by an asterisk to indicate correlations. In other cases, no 
information is given on the total. In those cases, the fraction $S_2$ is given 
in terms of the fraction $S_1$ in the table.

In case of GDR widths, some authors reduce their number of fit parameters by 
introducing a phenomenological relation between the widths and the centroids of
a two-component Lorentzian according to $\Gamma=c\,E^2$, where $c$ becomes the
fit parameter \cite{HB98}\@. In such cases, we have calculated the widths 
$\Gamma$ including their errors. However, the errors are again correlated and 
marked by an asterisk. In the case of two GDR centroids, some authors give the 
average $E_{\mathrm{ave}}=\frac{S_1\,E_1+S_2\,E_2}{S_1+S_2}$ and the ratio 
$E_2/E_1$\@. Sometimes this ratio is replaced by an average deformation $\beta$
which can be related to $E_2/E_1$ by either 
$\beta=\frac{2}{3}\sqrt{\frac{4\pi}{5}}\ln\frac{E_2}{E_1}$ (see, e.g., 
\cite{KS93}) or $\beta=\sqrt{\frac{4\pi}{5}}\frac{E_\perp/E_\parallel-1}
{E_\perp/2E_\parallel+0.8665}$ (see, e.g., \cite{HB98}), where $E_\perp$ and 
$E_\parallel$ denote the centroids of the GDR components due to oscillations 
perpendicular and parallel to the symmetry axis. In the case where the original
article did not mention which of the two formulas applies \cite{TH88}, the two 
formulas resulted in values for $E_1$ and $E_2$ within 0.2~MeV of each other, a
difference far less than the quoted statistical error. In general, however, we 
did not concern ourselves with deformations. Only when oblate deformation was 
established from, e.g., the asymmetry $a_2$, a '-' sign was added to the 
deformation parameter (in cases where the original article only provided the 
absolute value, see, e.g., \cite{GB88})\@.

In our treatment of errors, the first step was to add quadratically statistical
and systematical errors (when quoted separately)\@. Where a range of systematic
uncertainties is given, we adopted the center value of that range as a 
representative systematic uncertainty. When uncertainties are given in terms of
a $\mathrm{FWHM}$, it was converted by 
$\sigma={\mathrm{FWHM}}/\sqrt{8\,\ln 2}$\@. In general, rigorous error 
propagation was performed. However, no original work published the full 
covariance matrix for the fitted GDR parameters. Therefore, the derived errors 
are only representative of the true errors under the assumption that the 
originally fitted parameters are fully uncorrelated. In cases where we 
determined more parameters than were originally fitted, an asterisk denotes the
correlations which were introduced to the errors. In cases where GDR parameters
were held fixed during the fit, a little 'f' was added in the table instead of 
an error. Some works do not cite errors at all. In such cases we have made no 
attempt to estimate the errors. In a few cases where $E_1$ and $E_2$ were given
without errors, while $E_{\mathrm{ave}}$ was given with error, we could not 
find a good method to translate this error into errors of the individual 
values, and hence $E_1$ and $E_2$ remain without errors (see, e.g., 
\cite{GB88})\@. 

\subsection{Policy}

Although we have tried to avoid a true evaluation of the original articles, we 
have excluded some of them from the present compilation. Typically, where 
high-energy $\gamma$ rays from sources other than a compound nucleus were 
investigated (such as the mononucleus as in \cite{TG96} or the di-nuclear 
system as in \cite{CR95}), the resulting data are not used for the present 
compilation. In a different case, the data in question were heavily 
contaminated by pile-up events which lead to unphysical GDR parameters 
\cite{MS90}\@. In general, when different fits to the same data were performed 
(using different input parameters for the statistical-model calculation such as
the level-density formula, see, e.g., \cite{FS93}, or the nuclear viscosity in 
the case of an open fission channel, see, e.g., \cite{HB98}), we present all 
possible fits. On the other hand, when essentially the same data were presented
in a conference proceedings as well as in a subsequent refereed article with 
no or minimal differences in the fit parameters, we typically report only the 
results from the refereed article with few exceptions. Finally, when no GDR 
parameters were given (see, e.g., \cite{HM83,ZB97,MK05} or when we feel that 
the statistical-model description of the experiment is rather tentative 
\cite{NH81,SA95}, we made no attempt to fit the experimental spectrum 
ourselves, hence such data are not taken into account in the present 
compilation.

\subsection{Note on References}

When data from this compilation are cited, reference should also be made to the
original publication as well.

\ack

We would like to thank Chris Kawatsu and Mark Shevin for the initial literature
search. This work was supported by National Science Foundation Grant No.\ 
PHY01-10253\@.

\newpage

\section*{EXPLANATION OF TABLE}
\label{tableI}
\addcontentsline{toc}{section}{EXPLANATION OF TABLE}

\noindent\textbf{TABLE\@. Reaction and GDR parameters}

\begin{longtable}{lp{5.9in}}
No.&number of item\\
Ref.&reference\\
Pro.&projectile nucleus\\
Tar.&target nucleus\\
CN&compound nucleus as determined by the gating conditions\\
$E_{\mathrm{lab}}$&laboratory beam energy (MeV), sometimes given at the center 
of the target\\
$E_{\mathrm{ex}}$&excitation energy of the compound nucleus (MeV) as determined
by the gating conditions, sometimes averaged over the target thickness\\
$\langle I_i\rangle$&initial average angular momentum of the compound nucleus 
($\hbar$) as determined by the gating conditions, sometimes calculated by 
$\langle I_i\rangle=\frac{2}{3}l_0$ where $l_0$ is the maximum spin of the 
compound system calculated by the theories of Winther or Swiatecki\\
\end{longtable}

\newpage

\datatables

\setlength{\LTleft}{0pt}
\setlength{\LTright}{0pt} 
\setlength{\tabcolsep}{0.5\tabcolsep}

\begin{longtable}{p{6.725in}}
\caption[Hot GDR parameters]{Giant Electric Dipole Resonance parameters.}\\
\multicolumn{1}{c}{\small See page \protect\pageref{tableI} for Explanation of 
Tables.}\\\hline
\\[100pt]
\multicolumn{1}{c}{\footnotesize\it This space intentionally left blank}
\end{longtable}

\newpage

\footnotesize

\begin{tabular}{rcccccccccccccccc}
No.&Ref.&Pro.&Tar.&CN&$E_{\mathrm{lab}}$&$E_{\mathrm{ex}}$&$\langle I_i\rangle$
&$T_i$&l.d.&$A/a$&$\langle I_f\rangle$&$\mathrm{FWHM}$&$E_f$&$T_f$&Code&gate\\
  1&Kic90   &$^{ 12}$C &$^{         27} $Al      &$^{ 39}$K &     33.8&     40.0&{\it  7.7}&    & R & 8  &      8.0 &          &14.9&1.75                 & C  &                            \\
  2&Kic90   &$^{ 12}$C &$^{         27} $Al      &$^{ 39}$K &     48.2&     50.0&{\it 13.2}&    & R & 8  &     13.0 &          &14.3&1.71                 & C  &                            \\
  3&Kic90   &$^{ 12}$C &$^{         27} $Al      &$^{ 39}$K &     62.7&     60.0&{\it 16.5}&    & R & 8  &     16.0 &          &13.9&1.69                 & C  &                            \\
  4&Kic90   &$^{ 13}$C &$^{         27} $Al      &$^{ 40}$K &     39.7&     46.3&{\it  9.3}&    & R & 8  &      9.5 &          &17.8&1.88                 & C  &                            \\
  5&Kic90   &$^{ 15}$N &$^{         27} $Al      &$^{ 42}$Ca&     29.2&     40.2&{\it  7.7}&    & R & 8  &      8.0 &          &16.5&1.77                 & C  &                            \\
  6&Kic90   &$^{ 15}$N &$^{         27} $Al      &$^{ 42}$Ca&     44.7&     50.2&{\it 12.5}&    & R & 8  &     12.5 &          &16.7&1.78                 & C  &                            \\
  7&Kic90   &$^{ 18}$O &$^{         27} $Al      &$^{ 45}$Sc&     31.5&     42.0&{\it  7.7}&    & R & 8  &      8.0 &          &16.1&1.69                 & C  &                            \\
\end{tabular}

\newpage 

\begin{tabular}{rcccccccc}
No.&$S_1$&$E_1$&$\Gamma_1$&$S_2$&$E_2$&$\Gamma_2$&$\mathrm{FWHM}$&def.\\
  1&     0.89  $\pm$  {\it 0.17}          &     16.4   $\pm$  {\it 0.5}         &     10.5  $\pm$  {\it 1.0}     &                                      &                                      &                           &                       &                                               \\
  2&     0.77  $\pm$  {\it 0.13}          &     16.8   $\pm$  {\it 0.5}         &     11.6  $\pm$  {\it 1.1}     &                                      &                                      &                           &                       &                                               \\
  3&     0.86  $\pm$  {\it 0.14}          &     17.0   $\pm$  {\it 0.5}         &     13.4  $\pm$  {\it 1.1}     &                                      &                                      &                           &                       &                                               \\
  4&     1.33  $\pm$  {\it 0.22}          &     17.2   $\pm$  {\it 0.5}         &     12.1  $\pm$  {\it 1.1}     &                                      &                                      &                           &                       &                                               \\
  5&     1.19  $\pm$  {\it 0.20}          &     17.7   $\pm$  {\it 0.5}         &     11.6  $\pm$  {\it 1.0}     &                                      &                                      &                           &                       &                                               \\
  6&     1.17  $\pm$  {\it 0.19}          &     17.1   $\pm$  {\it 0.5}         &     11.9  $\pm$  {\it 1.0}     &                                      &                                      &                           &                       &                                               \\
\end{tabular}

\newpage

\begin{tabular}{rcccccccccccccccc}
No.&Ref.&Pro.&Tar.&CN&$E_{\mathrm{lab}}$&$E_{\mathrm{ex}}$&$\langle I_i\rangle$
&$T_i$&l.d.&$A/a$&$\langle I_f\rangle$&$\mathrm{FWHM}$&$E_f$&$T_f$&Code&gate\\
 58&Kic87   &$^{ 18}$O &$^{         45} $Sc      &$^{ 63}$Cu&     53.1&     61.6&     15.5 &    & R & 8  &     15.0 &          &27.1&1.86                 & C  &                            \\
 59&Kic87   &$^{ 12}$C &$^{         51} $V       &$^{ 63}$Cu&     64.7&     65.7&     20.0 &    & R & 8  &     20.5 &          &23.2&1.72                 & C  &                            \\
 60&Dre95   &$^{ 18}$O &$^{         45} $Sc      &$^{ 63}$Cu&     68  &     72.4&     23.3 &    & R & 8  &     19.6 &          &    &1.9                  & C  &                            \\
 61&Kic87   &$^{ 18}$O &$^{         45} $Sc      &$^{ 63}$Cu&     75.1&     77.4&     23.0 &    & R & 8  &     22.0 &          &29.2&1.93                 & C  &                            \\
 62&Dre95   &$^{ 18}$O &$^{         45} $Sc      &$^{ 63}$Cu&    105  &     99.1&     25.3 &    & R & 8  &     22.6 &          &    &2.2                  & C  &                            \\
 63&Dre95   &$^{ 18}$O &$^{         45} $Sc      &$^{ 63}$Cu&    125  &    113.0&     25.3 &    & R & 8  &     22.9 &          &    &2.3                  & C  &                            \\
 64&Dre95   &$^{ 18}$O &$^{         45} $Sc      &$^{ 63}$Cu&    149  &    130.1&     25.3 &    & R & 8  &     22.9 &          &    &2.5                  & C  &                            \\
\end{tabular}

\newpage 

\begin{tabular}{rcccccccc}
No.&$S_1$&$E_1$&$\Gamma_1$&$S_2$&$E_2$&$\Gamma_2$&$\mathrm{FWHM}$&def.\\
 58&     0.99  $\pm$       0.10           &     16.3   $\pm$       0.3          &      9.9  $\pm$       0.5      &                                      &                                      &                           &                       &                                               \\
 59&     1.07  $\pm$       0.10           &     17.0   $\pm$       0.3          &     10.5  $\pm$       0.5      &                                      &                                      &                           &                       &                                               \\
 60&     0.9   $\pm$       0.1            &     17.3   $\pm$       0.3          &     11.4  $\pm$       0.6      &                                      &                                      &                           &                       &                                               \\
 61&     0.90  $\pm$       0.10           &     16.4   $\pm$       0.3          &     10.6  $\pm$       0.6      &                                      &                                      &                           &                       &                                               \\
 62&     1.1   $\pm$       0.1            &     16.8   $\pm$       0.3          &     14.8  $\pm$       0.9      &                                      &                                      &                           &                       &                                               \\
 63&     1.1   $\pm$  {\it 0.2}           &     16.7   $\pm$  {\it 0.3}         &     12.7  $\pm$  {\it 2.3}     &                                      &                                      &                           &                       &                                               \\
\end{tabular}

\newpage

\begin{tabular}{rcccccccccccccccc}
No.&Ref.&Pro.&Tar.&CN&$E_{\mathrm{lab}}$&$E_{\mathrm{ex}}$&$\langle I_i\rangle$
&$T_i$&l.d.&$A/a$&$\langle I_f\rangle$&$\mathrm{FWHM}$&$E_f$&$T_f$&Code&gate\\
115&Bra95   &$^{ 48}$Ti&$^{         62} $Ni      &$^{110}$Sn&    223  &     92  &     40   &    &   & 8  &     45   &     24   &    &                     & C  &f$>$9                       \\
116&Bra95   &$^{ 48}$Ti&$^{         62} $Ni      &$^{110}$Sn&    223  &     92  &     40   &1.96&   & 8  &     44   &     16   &    &1.8  $\pm$ {\it 0.13}& C  &$\gamma$m                   \\
117&Bra95   &$^{ 48}$Ti&$^{         62} $Ni      &$^{110}$Sn&    223  &     92  &     40   &1.74&   & 8  &     54   &     14   &    &1.6  $\pm$ {\it 0.13}& C  &$\gamma$m                   \\
118&Cha87a  &$^{ 16}$O &$^{         94} $Mo      &$^{110}$Sn&    140  &    110.2&          &    &   & 9  &          &          &    &                     & C  &$\gamma$m                   \\
119&Bra89   &$^{ 40}$Ar&$^{         70} $Ge      &$^{110}$Sn&    400  &    230  &{\it 40}  &    &   &    &          &          &    &                     & C  &er                          \\
120&Gaa86   &$^{ 20}$Ne&$^{         91} $Zr      &$^{111}$Sn&    100  &     66.4&{\it 27}  &    &   & 8  &          &          &    &                     & C  &$\gamma$m                   \\
121&Gaa86   &$^{ 20}$Ne&$^{         91} $Zr      &$^{111}$Sn&    140  &     99.2&{\it 41}  &    &   & 8  &          &          &    &                     & C  &$\gamma$m                   \\
\end{tabular}

\newpage 

\begin{tabular}{rcccccccc}
No.&$S_1$&$E_1$&$\Gamma_1$&$S_2$&$E_2$&$\Gamma_2$&$\mathrm{FWHM}$&def.\\
115&     1.00               f             &     14.9   $\pm$       0.5          &     11.8  $\pm$       0.6      &                                      &                                      &                           &                       &                                               \\
116&     1.00               f             &     15.0   $\pm$       0.500        &     11.7  $\pm$       0.600    &                                      &                                      &                           &                       &                                               \\
117&     1.00               f             &     14.7   $\pm$       0.500        &     12.8  $\pm$       0.600    &                                      &                                      &                           &                       &                                               \\
118&     1.00               f             &     14.9   $\pm$       0.3          &      9.8  $\pm$       0.5      &                                      &                                      &                           &                       &                                               \\
119&     1.00               f             &     16     $\pm$       1            &     13    $\mathit{^{+1}_{-2}}$&                                      &                                      &                           &                       &                                               \\
120&     1.0   $\pm$       0.1            &     15.5                            &      7.5                       &                                      &                                      &                           &                       &                                               \\
\end{tabular}

\newpage

\begin{tabular}{rcccccccccccccccc}
No.&Ref.&Pro.&Tar.&CN&$E_{\mathrm{lab}}$&$E_{\mathrm{ex}}$&$\langle I_i\rangle$
&$T_i$&l.d.&$A/a$&$\langle I_f\rangle$&$\mathrm{FWHM}$&$E_f$&$T_f$&Code&gate\\
172&Kmi00   &$^{ 37}$Cl&$^{        110} $Pd      &$^{147}$Eu&    165  &     77.6&{\it 41}  &    & R &    &     44   &          &    &1.29                 & C  &f=5--30                     \\
173&Kmi00   &$^{ 37}$Cl&$^{        110} $Pd      &$^{147}$Eu&    165  &     77.6&{\it 41}  &    & R &    &     49   &          &    &1.24                 & C  &f=8--30                     \\
174&Kmi00   &$^{ 37}$Cl&$^{        110} $Pd      &$^{147}$Eu&    170  &     81.4&{\it 43.0}&    & R &    &     39   &          &    &1.36                 & C  &f=5--7                      \\
175&Kmi00   &$^{ 37}$Cl&$^{        110} $Pd      &$^{147}$Eu&    170  &     81.4&{\it 43.0}&    & R &    &     45   &          &    &1.31                 & C  &f=5--30                     \\
176&Kmi00   &$^{ 37}$Cl&$^{        110} $Pd      &$^{147}$Eu&    170  &     81.4&{\it 43.0}&    & R &    &     50   &          &    &1.28                 & C  &f=8--30                     \\
177&Kmi00   &$^{ 37}$Cl&$^{        110} $Pd      &$^{147}$Eu&    170  &     81.4&{\it 43.0}&    & R &    &     39   &          &    &1.36                 & C  &f=5--7                      \\
178&Kmi00   &$^{ 37}$Cl&$^{        110} $Pd      &$^{147}$Eu&    170  &     81.4&{\it 43.0}&    & R &    &     45   &          &    &1.31                 & C  &f=5--30                     \\
\end{tabular}

\newpage 

\begin{tabular}{rcccccccc}
No.&$S_1$&$E_1$&$\Gamma_1$&$S_2$&$E_2$&$\Gamma_2$&$\mathrm{FWHM}$&def.\\
172&     0.42  $\pm$       0.02           &     13.0   $\pm$       0.1          &      5.8  $\pm$       0.1      &     0.43  $\pm$       0.02           &     16.7   $\pm$       0.1           &      7.4  $\pm$      0.2  &                       &         $\pm$0.27  $\pm$       0.01           \\
173&     0.54  $\pm$       0.02           &     12.9   $\pm$       0.1          &      5.8  $\pm$       0.1      &     0.48  $\pm$       0.03           &     17.0   $\pm$       0.1           &      7.1  $\pm$      0.4  &                       &         $\pm$0.3   $\pm$       0.01           \\
174&     0.91  $\pm$       0.03           &     14.2   $\pm$       0.1          &      8.8  $\pm$       0.4      &                                      &                                      &                           &                       &                                               \\
175&     0.98  $\pm$       0.03           &     14.1   $\pm$       0.1          &      8.9  $\pm$       0.3      &                                      &                                      &                           &                       &                                               \\
176&     1.01  $\pm$       0.04           &     14.0   $\pm$       0.1          &      8.8  $\pm$       0.3      &                                      &                                      &                           &                       &                                               \\
177&     0.43  $\pm$       0.02           &     12.9   $\pm$       0.1          &      5.9  $\pm$       0.1      &     0.41  $\pm$       0.03           &     16.8   $\pm$       0.1           &      7.3  $\pm$      0.4  &                       &         $\pm$0.29  $\pm$       0.01           \\
\end{tabular}

\newpage

\begin{tabular}{rcccccccccccccccc}
No.&Ref.&Pro.&Tar.&CN&$E_{\mathrm{lab}}$&$E_{\mathrm{ex}}$&$\langle I_i\rangle$
&$T_i$&l.d.&$A/a$&$\langle I_f\rangle$&$\mathrm{FWHM}$&$E_f$&$T_f$&Code&gate\\
229&Hof98   &$^{ 40}$Ar&$^{        116} $Cd      &$^{156}$Dy&    242.4&    124.0&{\it 70}  &    &   &10  &     89   &          &    &                     & C  &ff                          \\
230&Hof98   &$^{ 40}$Ar&$^{        116} $Cd      &$^{156}$Dy&    242.4&    124.0&{\it 70}  &    &   &10  &     89   &          &    &                     & C  &ff                          \\
231&Hof98   &$^{ 40}$Ar&$^{        116} $Cd      &$^{156}$Dy&    242.4&    124.0&{\it 70}  &    &   &10  &     89   &          &    &                     & C  &ff                          \\
232&Hof98   &$^{ 40}$Ar&$^{        116} $Cd      &$^{156}$Dy&    242.4&    124.0&{\it 70}  &    &   &10  &     89   &          &    &                     & C  &ff                          \\
233&Vie96   &$^{ 64}$Ni&$^{         92} $Zr      &$^{154}$Er&    241  &         &          &    &   & 8  &{\it 17.5}&{\it 25}  &    &                     & C  &d$\gamma$t                  \\
234&Vie97   &$^{ 64}$Ni&$^{         92} $Zr      &$^{154}$Er&    241  &         &{\it 31}  &    &   &    &          &          &    &                     & C  &d$\gamma$t                  \\
235&Vie97   &$^{ 64}$Ni&$^{         92} $Zr      &$^{155}$Er&    241  &         &{\it 31}  &    &   &    &          &          &    &                     & C  &d$\gamma$t                  \\
\end{tabular}

\newpage 

\begin{tabular}{rcccccccc}
No.&$S_1$&$E_1$&$\Gamma_1$&$S_2$&$E_2$&$\Gamma_2$&$\mathrm{FWHM}$&def.\\
229&{\it 0.21} $\pm$  {\it 0.01}          &{\it 11.20} $\pm$  {\it 0.14}        & {\it 4.6} $\pm$  {\it 1.5*}    &{\it 0.84} $\pm$  {\it 0.02}          &{\it 16.21} $\pm$  {\it 0.06}         & {\it 9.7} $\pm$ {\it 3.2*}&                       &              0.446 $\pm$       0.018          \\
230&{\it 0.24} $\pm$  {\it 0.03}          &{\it 11.03} $\pm$  {\it 0.14}        & {\it 3.8} $\pm$  {\it 0.5*}    &{\it 0.91} $\pm$  {\it 0.09}          &{\it 15.97} $\pm$  {\it 0.13}         & {\it 7.9} $\pm$ {\it 1.0*}&                       &              0.446 $\pm$       0.015          \\
231&{\it 0.21} $\pm$  {\it 0.04}          &{\it 11.40} $\pm$  {\it 0.06}        & {\it 5.0} $\pm$  {\it 0.2*}    &{\it 0.86} $\pm$  {\it 0.14}          &{\it 16.14} $\pm$  {\it 0.02}         & {\it 9.9} $\pm$ {\it 0.4*}&                       &              0.419 $\pm$       0.008          \\
232&{\it 0.21} $\pm$  {\it 0.01}          &{\it 11.94} $\pm$  {\it 0.10}        & {\it 6.7} $\pm$  {\it 0.2*}    &{\it 0.80} $\pm$  {\it 0.01}          &{\it 16.01} $\pm$  {\it 0.05}         &{\it 12.0} $\pm$ {\it 0.2*}&                       &              0.352 $\pm$       0.012          \\
233&     0.90  $\pm$       0.30           &     15.1   $\pm$       0.6          &      9.0  $\pm$       1.0      &                                      &                                      &                           &                       &                                               \\
234&     1                  f             &     15                  f           &      7                 f       &                                      &                                      &                           &                       &                                               \\
\end{tabular}

\newpage

\begin{tabular}{rcccccccccccccccc}
No.&Ref.&Pro.&Tar.&CN&$E_{\mathrm{lab}}$&$E_{\mathrm{ex}}$&$\langle I_i\rangle$
&$T_i$&l.d.&$A/a$&$\langle I_f\rangle$&$\mathrm{FWHM}$&$E_f$&$T_f$&Code&gate\\
286&Cha96   &$^{ 16}$O &$^{        181} $Ta      &$^{197}$Tl&     94.1&     60.1&{\it 24}  &    & I &    &{\it 28}  &          &    &                     & C  &f=7--14                     \\
287&Cha96   &$^{ 16}$O &$^{        181} $Ta      &$^{197}$Tl&    106.5&     71.6&{\it 33}  &    & I &    &          &          &    &                     & C  &$\gamma$c                   \\
288&Tho88   &$^{ 19}$F &$^{        181} $Ta      &$^{200}$Pb&     93  &     57.6&     14   &    &   & 8  &{\it  7}  &{\it 14}  &    &0.9                  & C  &                            \\
289&        &          &                         &          &         &         &          &    &   & 8  &{\it 18.5}&{\it  7}  &    &0.9                  & C  &                            \\
290&Gel00   &$^{ 19}$F &$^{        181} $Ta      &$^{200}$Pb&    101  &     65  &{\it 21}  &    &   & 8  &          &          &    &                     & C  &f$\ge$6                     \\
291&Cha87b  &$^{ 16}$O &$^{\mathrm{nat}}$W       &        Pb&    100  &     65.7&     25   &    &   & 8  &          &          &    &1.1                  & C  &$\gamma$m                   \\
292&Cha87b  &$^{ 19}$F &$^{        181} $Ta      &$^{200}$Pb&    105  &     68.5&          &    &   & 8  &          &          &    &1.1                  & C  &$\gamma$m                   \\
\end{tabular}

\newpage 

\begin{tabular}{rcccccccc}
No.&$S_1$&$E_1$&$\Gamma_1$&$S_2$&$E_2$&$\Gamma_2$&$\mathrm{FWHM}$&def.\\
286&{\it 0.33}         {\it f}            &     11.3   $\pm$       0.2          &      3.5  $\pm$       0.5      &{\it 0.67}         {\it f}            &     14.4   $\pm$       0.7           &      5.5  $\pm$      0.7  &                       &              0.34  $\mathit{^{+0.01}_{-0.02}}$\\
287&{\it 0.33}         {\it f}            &     11.4   $\pm$       0.2          &      5.1  $\pm$       0.5      &{\it 0.67}         {\it f}            &     14.8   $\pm$       0.7           &      8.0  $\pm$      0.7  &                       &              0.31  $\mathit{^{+0.03}_{-0.02}}$\\
288&                                      &     14.0   $\pm$       0.2          &      5.0  $\pm$       0.3      &                                      &                                      &                           &                       &                                               \\
289&                                      &     11.7   $\pm$       0.2          &      3.5  $\pm$       0.3      &                                      &     16.5   $\pm$       0.4           &      5.5  $\pm$      0.5  &                       &              0.41  $\pm$       0.04           \\
290&{\it 0.33} $\pm$  {\it 0.06}          &     11.7   $\pm$       0.3          &      3.9  $\pm$       0.4      &{\it 0.67} $\pm$  {\it 0.08}          &     15.0   $\pm$       0.5           &      4.8  $\pm$      0.5  &                       &              0.29  $\pm$       0.08           \\
291&                                      &     11.9   $\pm$       0.1          &      3.8  $\pm$       0.2      &           $2S_1$       f             &     15.4   $\pm$       0.2           &      5.3  $\pm$      0.3  &                       &              0.31  $\pm$       0.02           \\
\end{tabular}

\newpage

\begin{theDTbibliography}{00000000}

\bibitem[Ata90]{Ata90}
A. Ata\c{c}, J. J. Gaardh{\o}je, B. Herskind, Y. Iwata, S. Ogaza,
\newblock Phys.\ Lett.\ B {\bf 252}, 545 (1990).

\bibitem[Bal94]{Bal94}
A. Ba{\l}anda, J. C. S. Bacelar, E. B\v{e}t\'{a}k, J. A. Bordewijk, 
A. Krasznahorka, H. van der Ploeg, R. H. Siemssen, H. W. Wilschut, 
A. van der Woude,
\newblock Nucl.\ Phys.\ A {\bf 575}, 348 (1994).

\bibitem[Bau98]{Bau98}
T. Baumann, E. Ramakrishnan, A. Azhari, J. R. Beene, R. J. Charity,
{\it et al.},
\newblock Nucl.\ Phys.\ A {\bf 635}, 428 (1998);
E. Ramakrishnan, T. Baumann, A. Azhari, R.A. Kryger, R. Pfaff, {\it et al.},
\newblock Phys.\ Rev.\ Lett.\ {\bf 76}, 2025 (1996).

\bibitem[Bra05]{Bra05}
A. Bracco, O. Wieland
\newblock AIP Conf.\ Proc.\ {\bf 802}, 175 (2005).

\bibitem[Bra89]{Bra89}
A. Bracco, J. J. Gaardh{\o}je, A. M. Bruce, J. D. Garrett, B. Herskind,
{\it et al.},
\newblock Phys.\ Rev.\ Lett.\ {\bf 62}, 2080 (1989).

\bibitem[Bra95]{Bra95}
A. Bracco, F. Camera, M. Mattiuzzi, B. Million, M. Pignanelli,
J. J. Gaardh{\o}je, A. Maj, T. Rams{\o}y, T. Tveter, Z. \'{Z}elazny,
\newblock Phys.\ Rev.\ Lett.\ {\bf 74}, 3748 (1995);
A. Bracco, F. Camera,
\newblock Z. Phys.\ A {\bf 349}, 213 (1994).

\bibitem[Bra96]{Bra96}
A. Bracco, F. Camera, M. Mattiuzzi, D. Gnaccolini, B. Million,
J. J. Gaardh{\o}je, A. Maj, T. Tveter,
\newblock Nucl.\ Phys.\ A {\bf 599}, 83c (1996).

\bibitem[Bru88]{Bru88}
A. M. Bruce, J. J. Gaardh{\o}je, B. Herskind, R. Chapman, J. C. Lisle,
F. Khazaie, J. N. Mo, P. J. Twin,
\newblock Phys.\ Lett.\ B {\bf 215}, 237 (1988).

\bibitem[But90]{But90}
R. Butsch, M. Thoennessen, D. R. Chakrabarty, M. G. Herman, P. Paul,
\newblock Phys.\ Rev.\ C {\bf 41}, 1530 (1990).

\bibitem[Cam03]{Cam03}
F. Camera, A. Bracco, V. Nanal, M. P. Carpenter, F. Della Vedova, 
{\it et al.},
\newblock Phys.\ Lett.\ B {\bf 560}, 155 (2003).

\bibitem[Cam99]{Cam99}
F. Camera, A. Bracco, S. Leoni, B. Million, M. Mattiuzzi, {\it et al.},
\newblock Phys.\ Rev.\ C {\bf 60}, 014306 (1999);
F. Camera, A. Bracco, G. Colombo, S. Leoni, B. Million, {\it et al.},
\newblock Nucl.\ Phys.\ A {\bf 649}, 115c (1999).

\bibitem[Cha87a]{Cha87a}
D. R. Chakrabarty, S. Sen, M. Thoennessen, N. Alamanos, P. Paul, R. Schicker,
J. Stachel, J. J. Gaardhoje,
\newblock Phys.\ Rev.\ C {\bf 36}, 1886 (1987).

\bibitem[Cha87b]{Cha87b}
D. R. Chakrabarty, M. Thoennessen, N. Alamanos, P. Paul, S. Sen,
\newblock Phys.\ Rev.\ Lett.\ {\bf 58}, 1092 (1987);
Dipak R. Chakrabarty,
\newblock Nucl.\ Phys.\ A {\bf 482}, 81c (1988).

\bibitem[Cha88]{Cha88}
D. R. Chakrabarty, M. Thoennessen, S. Sen, P. Paul, R. Butsch, M. G. Herman,
\newblock Phys.\ Rev.\ C {\bf 37}, 1437 (1988).

\bibitem[Cha96]{Cha96}
D. R. Chakrabarty, V. M. Datar, R. K. Choudhury, B. K. Nayak, Y. K. Agarwal,
C. V. K. Baba, M. K. Sharan,
\newblock Phys.\ Rev.\ C {\bf 53}, 2739 (1996).

\bibitem[Di\'{o}00a]{Dio00a}
I. Di\'{o}szegi, N. P. Shaw, A. Bracco, F. Camera, S. Tettoni, M. Mattiuzzi,
P. Paul,
\newblock Phys.\ Rev.\ C {\bf 63}, 014611 (2000).

\bibitem[Di\'{o}00b]{Dio00b}
I. Di\'{o}szegi, N. P. Shaw, I. Mazumdar, A. Hatzikoutelis, P. Paul,
\newblock Phys.\ Rev.\ C {\bf 61}, 024613 (2000).

\bibitem[Di\'{o}01]{Dio01}
I. Di\'{o}szegi, I. Mazumdar, N. P. Shaw, P. Paul,
\newblock Phys.\ Rev.\ C {\bf 63}, 047601 (2001).

\bibitem[Di\'{o}92]{Dio92}
I. Di\'{o}szegi, D. J. Hofman, C. P. Montoya, S. Schadmand, P. Paul,
\newblock Phys.\ Rev.\ C {\bf 46}, 627 (1992);
R. Butsch, D. J. Hofman, C. P. Montoya, P. Paul, M. Thoennessen,
\newblock Phys.\ Rev.\ C {\bf 44}, 1515 (1991).

\bibitem[Dre95]{Dre95}
Z. M. Drebi, K. A. Snover, A. W. Charlop, M. S. Kaplan, D. P. Wells, D. Ye,
Y. Alhassid,
\newblock Phys.\ Rev.\ C {\bf 52}, 578 (1995).

\bibitem[End92]{End92}
G. Enders, F. D. Berg, K. Hagel, W. K\"{u}hn, V. Metag, {\it el al.},
\newblock Phys.\ Rev.\ Lett.\ {\bf 69}, 249 (1992).

\bibitem[Fel93]{Fel93}
G. Feldman, K. A. Snover, J. A. Behr, C. A. Gossett, J. H. Gundlach,
M. Kici\'{n}ska-Habior,
\newblock Phys.\ Rev.\ C {\bf 47}, 1436 (1993).

\bibitem[Fli91]{Fli91}
S. Flibotte, B. Haas, P. Taras, H. R. Andrews, D. C. Radford, D. Ward,
\newblock Nucl.\ Phys.\ A {\bf 531}, 205 (1991).

\bibitem[Fli96]{Fli96}
S. Flibotte, M. Cromaz, J. DeGraaf, T. E. Drake, A. Galindo-Uribarri,
{\it et al.},
\newblock Phys.\ Rev.\ C {\bf 53}, R533 (1996).

\bibitem[Gaa84]{Gaa84}
J. J. Gaardh{\o}je, C. Ellegaard, B. Herskind, S. G. Steadman,
\newblock Phys.\ Rev.\ Lett.\ {\bf 53}, 148 (1984).

\bibitem[Gaa86]{Gaa86}
J. J. Gaardh{\o}je, C. Ellegaard, B. Herskind, R. M. Diamond,
M. A. Deleplanque, G. Dines, A. O. Macchiavelli, F. S. Stephens,
\newblock Phys.\ Rev.\ Lett.\ {\bf 56}, 1783 (1986).

\bibitem[Gaa88]{Gaa88}
J. J. Gaardh{\o}je, A. M. Bruce, B. Herskind,
\newblock Nucl.\ Phys.\ A {\bf 482}, 121c (1988).

\bibitem[Gar83]{Gar83}	
E. F. Garman, K. A. Snover, S. H. Chew, S. K. B. Hesmondhalgh, W. N. Catford,
P. M. Walker,
\newblock Phys.\ Rev.\ C {\bf 28}, 2554 (1983); 
A. Lazzarini, D. Habs, W. Hennerici, R. Kroth, J. Schirmer, and V. Metag, 
\newblock Phys.\ Rev.\ Lett.\ {\bf 53}, 1045 (1984).

\bibitem[Gel00]{Gel00}
N. Gelli, F. Lucarelli, M. Cinausero, E. Fioretto, G. Prete,
{\it et al.},
\newblock Eur.\ Phys.\ J. A {\bf 7}, 361 (2000).

\bibitem[Gos85]{Gos85}
C. A. Gossett, K. A. Snover, J. A. Behr, G. Feldman, J. L. Osborne,
\newblock Phys.\ Rev.\ Lett.\ {\bf 54}, 1486 (1985).

\bibitem[Gun90]{Gun90}
J. H. Gundlach, K. A. Snover, J. A. Behr, C. A. Gossett,
M. Kicinska-Habior, K. T. Lesko,
\newblock Phys.\ Rev.\ Lett.\ {\bf 65}, 2523 (1990).

\bibitem[Haa83]{Haa83}
B. Haas, D. C. Radford, F. A. Beck, T. Byrski, C. Gehringer, J. C. Merdinger,
A. Nourredine, Y. Schutz, J. P. Vivien,
\newblock Phys.\ Lett.\ {\bf 120B}, 79 (1983).

\bibitem[Hec03]{Hec03}
P. Heckman, D. Bazin, J. R. Beene, Y. Blumenfeld, M. J. Chromik,
{\it et al.},
\newblock Phys.\ Lett.\ B {\bf 555}, 43 (2003).

\bibitem[Hof94]{Hof94}
D. J. Hofman, B. B. Back, I. Di\'{o}szegi, C. P. Montoya, S. Schadmand,
R. Varma, P. Paul,
\newblock Phys.\ Rev.\ Lett.\ {\bf 72}, 470 (1994).

\bibitem[Hof98]{Hof98}
G. van 't Hof, J. C. S. Bacelar, I. Di\'{o}szegi, M. N. Harakeh,
W. H. A. Hesselink, N. Kalantar-Nayestanaki, A. Kugler, H. van der Ploeg,
A. J. M. Plompen, J. P. S. van Schagen,
\newblock Nucl.\ Phys.\ A {\bf 638}, 613 (1998).

\bibitem[Kel97]{Kel97}
M. P. Kelly, J. F. Liang, A. A. Sonzogni, K. A. Snover, J. P. S. van Schagen,
J. P. Lestone,
\newblock Phys.\ Rev.\ C {\bf 56}, 3201 (1997).

\bibitem[Kel99]{Kel99}
M. P. Kelly, K. A. Snover, J. P. S. van Schagen, M. Kici\'{n}ska-Habior,
Z. Trznadel,
\newblock Phys.\ Rev.\ Lett.\ {\bf 82}, 3404 (1999).

\bibitem[Kic87]{Kic87}
M. Kici\'{n}ska-Habior, K. A. Snover, C. A. Gossett, J. A. Behr, G. Feldman,
H. K. Glatzel, J. H. Gundlach, E. F. Garman,
\newblock Phys.\ Rev.\ C {\bf 36}, 612 (1987).

\bibitem[Kic90]{Kic90}
M. Kici\'{n}ska-Habior, K. A. Snover, J. A. Behr, G. Feldman, C. A. Gossett,
J. H. Gundlach,
\newblock Phys.\ Rev.\ C {\bf 41}, 2075 (1990).

\bibitem[Kic92]{Kic92}
M. Kici\'{n}ska-Habior, K. A. Snover, J. A. Behr, C. A. Gossett, 
J. H. Gundlach, G. Feldman,
\newblock Phys.\ Rev.\ C {\bf 45}, 569 (1992).

\bibitem[Kic93]{Kic93}
M. Kici\'{n}ska-Habior, K. A. Snover, J. A. Behr, C. A. Gossett, Y. Alhassid,
N. Whelan,
\newblock Phys.\ Lett.\ B {\bf 308}, 225 (1993).

\bibitem[Kmi00]{Kmi00}
M. Kmiecik, A. Maj, A. Bracco, F. Camera, M. Casanova, S. Leoni, B. Million,
B. Herskind, R. A. Bark, W. E. Ormand,
\newblock Nucl.\ Phys.\ A {\bf 674}, 29 (2000);
A. Bracco, F. Camera, S. Leoni,
\newblock Nucl.\ Phys.\ A {\bf 682}, 449c (2001);
A. Bracco, F. Camera, S. Leoni, B. Million, A. Maj, M. Kmiecik,
\newblock Nucl.\ Phys.\ A {\bf 687}, 237c (2001).

\bibitem[Kmi01]{Kmi01}
M. Kmiecik, A. Maj, A. Bracco, F. Camera, B. Million, O. Wieland,
\newblock Eur.\ Phys.\ J. A {\bf 12}, 5 (2001).

\bibitem[Kr\'{o}92]{Kro92}
W. Kr\'{o}las, A. Maj, P. Bednarczyk, B. Fornal, W. M\c{e}czy\'{n}ski,
J. Stycze\'{n}, M. Zi\c{e}bli\'{n}ski,
\newblock Z. Phys.\ A {\bf 344}, 145 (1992).

\bibitem[Luc96]{Luc96}
F. Lucarelli, N. Gelli, P. Blasi, M. Cinausero, E. Fioretto,
{\it et al.},
\newblock Z. Phys.\ A {\bf 355}, 35 (1996).

\bibitem[Mat95]{Mat95}
M. Mattiuzzi, A. Bracco, F. Camera, B. Million, M. Pignanelli,
J. J. Gaardh{\o}je, A. Maj, T. Rams{\o}y, T. Tveter, Z. \'{Z}elazny,
\newblock Phys.\ Lett.\ B {\bf 364}, 13 (1995).

\bibitem[Mat97]{Mat97}
M. Mattiuzzi, A. Bracco, F. Camera, W. E. Ormand, J. J. Gaardh{\o}je,
A. Maj, B. Million, M. Pignanelli, T. Tveter,
\newblock Nucl.\ Phys.\ A {\bf 612}, 262 (1997).

\bibitem[Nan99]{Nan99}
V. Nanal, B. B. Back, D. J. Hofman, G. Hackman, D. Ackermann, {\it et al.},
\newblock Nucl.\ Phys.\ A {\bf 649}, 153c (1999).

\bibitem[Noo92]{Noo92}
R. F. Noorman, J. C. Bacelar, M. N. Harakeh, W. H. A. Hesselink, H. J. Hofmann,
{\it et al.},
\newblock Phys.\ Lett.\ B {\bf 292}, 257 (1992).

\bibitem[Noo94]{Noo94}
R. F. Noorman, J. C. Bacelar, M. N. Harakeh, W. H. A. Hesselink, H. J. Hofmann,
{\it et al.},
\newblock Nucl.\ Phys.\ A {\bf 574}, 501 (1994).

\bibitem[Rat03]{Rat03}
S. K. Rathi, D. R. Chakrabarty, V. M. Datar, Suresh Kumar, E. T. Mirgule, 
A. Mitra, V. Nanal, H. H. Oza,
\newblock Phys.\ Rev.\ C {\bf 67}, 024603 (2003);
D. R. Chakrabarty,
\newblock Nucl.\ Phys.\ A {\bf 687}, 184c (2001).

\bibitem[Sch93]{Sch93}
J. P. S. van Schagen, Y. Alhassid, J. C. Bacelar, B. Bush, M. N. Harakeh,
{\it et al.},
\newblock Phys.\ Lett.\ B {\bf 308}, 231 (1993).

\bibitem[Sch95]{Sch95}
J. P. S. van Schagen, Y. Alhassid, J. C. S. Bacelar, B. Bush, M. N. Harakeh,
{\it et al.},
\newblock Phys.\ Lett.\ B {\bf 343}, 64 (1995).

\bibitem[Sch96]{Sch96}
S. Schadmand, R. Varma, S. R. Banerjee, B. B. Back, D. J. Hofman, 
C. P. Montoya, P. Paul,
\newblock Nucl.\ Phys.\ A {\bf 599}, 89c (1996).

\bibitem[Sha00]{Sha00}
N. P. Shaw, I. Di\'{o}szegi, I. Mazumdar, A. Buda, C. R. Morton, {\it et al.},
\newblock Phys.\ Rev.\ C {\bf 61}, 044612 (2000).

\bibitem[Sto89a]{Sto89a}
A. Stolk, A. Balanda, M. N. Harakeh, W. H. A. Hesselink, J. Penninga,
H. Rijneveld,
\newblock Nucl.\ Phys.\ A {\bf 505}, 241 (1989).

\bibitem[Sto89b]{Sto89b}
A. Stolk, M. N. Harakeh, W. H. A. Hesselink, H. J. Hofmann, R. F. Noorman,
J. P. S. van Schagen, Z. Sujkowski, H. Verheul, M. J. A. de Voigt,
D. J. P. Witte,
\newblock Phys.\ Rev.\ C {\bf 40}, R2454 (1989).

\bibitem[Thi88]{Thi88}
P. Thirolf, D. Habs, D. Schwalm, R. D. Fischer, V. Metag,
\newblock Nucl.\ Phys.\ A {\bf 482}, 93c (1988).

\bibitem[Tho87]{Tho87}
M. Thoennessen, D. R. Chakrabarty, M. G. Herman, R. Butsch, P. Paul,
\newblock Phys.\ Rev.\ Lett.\ {\bf 59}, 2860 (1987).

\bibitem[Tho88]{Tho88}
M. Thoennessen, D. R. Chakrabarty, R. Butsch, M. G. Herman, P. Paul, S. Sen,
\newblock Phys.\ Rev.\ C {\bf 37}, 1762 (1988).

\bibitem[Tho92]{Tho92}
M. Thoennessen, J. R. Beene, F. E. Bertrand, C. Baktash, M. L. Halbert,
{\it et al.},
\newblock Phys.\ Lett.\ B {\bf 282}, 288 (1992).

\bibitem[Tho93]{Tho93}
M. Thoennessen, J. R. Beene, F. E. Bertrand, C. Baktash, M. L. Halbert,
D. J. Horen, D. G. Sarantites, W. Spang, D. Stracener,
\newblock Phys.\ Rev.\ Lett.\ {\bf 70}, 4055 (1993).

\bibitem[Tho95]{Tho95}
M. Thoennessen, E. Ramakrishnan, J. R. Beene, F. E. Bertrand, M. L. Halbert,
D. J. Horen, P. E. Mueller, R. L. Varner,
\newblock Phys.\ Rev.\ C {\bf 51}, 3148 (1995).

\bibitem[Vie89]{Vie89}
G. Viesti, M. Anghinolfi, P. F. Bortignon, P. Corvisiero, L. Djilavian,
{\it et al.},
\newblock Phys.\ Rev.\ C {\bf 40}, R1570 (1989).

\bibitem[Vie96]{Vie96}
G. Viesti, M. Lops, B. Fornal, D. Bazzacco, P. F. Bortignon, {\it et al.},
\newblock Nucl.\ Phys.\ A {\bf 604}, 81 (1996);
M. Cinausero, D. Bazzacco, P. F. Bortignon, G. De Angelis, D. Fabris, {\it et 
al.},
\newblock Nucl.\ Phys.\ A {\bf 599}, 111c (1996).

\bibitem[Vie97]{Vie97}
G. Viesti, B. Fornal, M. Cinausero,
\newblock Phys.\ Rev.\ C {\bf 55}, 1594 (1997).

\bibitem[Voj89]{Voj89}
R. J. Vojtech, R. Butsch, V. M. Datar, M. G. Herman, R. L. McGrath, P. Paul,
M. Thoennessen,
\newblock Phys.\ Rev.\ C {\bf 40}, R2441 (1989).

\end{theDTbibliography}

\end{document}